\documentclass[12pt,preprint]{aastex}
\usepackage{longtable}
\usepackage{color}
\everymath{\rm}
\everydisplay{\sf}

\begin{document}

\title{	Abundances of PNe in the Outer Disk of M31\altaffilmark{1}}

\author{Karen B. Kwitter}
\affil{Department of Astronomy, Williams College, Williamstown, MA 01267, USA}
\email{kkwitter@williams.edu}

\author{Emma M. M. Lehman}
\affil{Department of Astronomy, Williams College, Williamstown, MA 01267, USA}
\email{emmalehman@gmail.com}

\author{Bruce Balick}
\affil{Department of Astronomy, University of Washington, Seattle, WA 98195, USA}
\email{balick@astro.washington.edu}

\and

\author{R.B.C. Henry}
\affil{H.L. Dodge Department of Physics \& Astronomy, University of Oklahoma, Norman, OK 73019, USA}
\email{rhenry@ou.edu}

\altaffiltext{1}{Partially based on observations obtained with the Apache Point Observatory 3.5-meter telescope, which is owned and operated by the Astrophysical Research Consortium; and on observations obtained at the Gemini Observatory, which is operated by the Association of Universities for Research in Astronomy, Inc., under a cooperative agreement with the NSF on behalf of the Gemini partnership: the National Science Foundation (United 
States), the Science and Technology Facilities Council (United Kingdom), the 
National Research Council (Canada), CONICYT (Chile), the Australian Research Council (Australia), 
Ministerio da Ciencia e Tecnologia (Brazil) 
and Ministerio de Ciencia, Tecnolog\'{i}a e Innovaci\'{o}n Productiva (Argentina).}

\begin{abstract}

We present spectroscopic observations and chemical abundances of 16 planetary nebulae (PNe) in the outer disk of M31. The [O~III] $\lambda$4363 line is detected in all objects, allowing a direct measurement of the nebular temperature essential for accurate abundance determinations. Our results show that the abundances in these M31 PNe display the same correlations and general behaviors as Type~II PNe in the Milky Way Galaxy. We also calculate photoionization models to derive estimates of central star properties. From these we infer that our sample PNe, all near the bright-end cutoff of the Planetary Nebula Luminosity Function, originated from stars near 2 M$_{\odot}$. Finally, under the assumption that these PNe are located in M31's disk, we plot the oxygen abundance gradient, which appears shallower than the gradient in the Milky Way.
\end{abstract}


\section{Introduction}
H~II regions and planetary nebulae (PNe) are useful probes of the past and present chemical composition of the interstellar medium.  The $\alpha$-element abundances of an ensemble of H~II regions provide a snapshot of the current element content and radial gradients. (Some $\alpha$-elements are also detectable in B supergiants [e.g. \citet{T02}], though their galactocentric radius range is more restricted.) On the other hand, leaving carbon aside, the $\alpha$-element abundances found in a planetary nebula reflect those in the ISM at the time that the progenitor star formed.\footnote {According to stellar evolution theory, the central stars of PNe do not generally enrich the $\alpha$ elements in their atmospheres, with minor exceptions to be discussed later.}   Comparisons of H~II and PN abundances at various values of galactocentric radius, $R_{gal}$, represent the only means of measuring cumulative and historical $\alpha$-element production, providing a glimpse into the chemical histories and production processes in the disks of  various spiral galaxies. 

Furthermore, plots of $\alpha$-element abundances against one another constrain the production processes and the relative yields of each element.  In this manner the $\alpha$-element ratios can reflect the relative importance of the different $\alpha$-element enrichment rates by supernovae and Wolf-Rayet stars.  A comparison of such plots at different values of $R_{gal}$ may yield individual chemical ``fingerprints'' for galaxies and the subunits from which they may have been assembled. 

In previous papers Henry et al. (2004; HKB04) and Henry et al. (2010; H10) examined these questions for a sample of H~II regions and PNe in the solar region of the Milky Way (MW).  The results showed that halo and disk PNe are distinctive in some of their $\alpha$-element ratios.  Also, within the disk, log plots of various $\alpha$-elements against O/H show linear trends with slopes near unity, suggesting that although PN abundances may differ from one nebula to another, the process(es) that enriched the elements are the same. This is profoundly interesting since the progenitor stars of nearby PNe have wide-ranging initial masses, compositions, ages and birth sites. 

H10 confirmed that the O/H ratio shows a measurable slope as a function of $R_{gal}$ in the MW. Importantly, the slopes of log (O/H) vs. $R_{gal}$ for the PNe and the H~II regions are consistent if not congruous. H10 also found that the scatter of PNe values from the gradient trend is often substantially larger than the measurement uncertainties and the scatter in results for H~II regions. Estimates of $R_{gal}$ in the MW are notoriously uncertain since various distance-measurement methods often disagree, particularly at large $R_{gal}$. Some of the scatter in the O/H gradient found by H10  is undoubtedly the result of the uncertainties in $R_{gal}$. (Interestingly, they also concluded that {\it intrinsic} abundance variations could contribute significantly to the observed scatter, as we mention in \S4.3.) However, the distance-determination problem disappears in M31, where all PNe have readily measurable values of $R_{gal}$, and where the key faint diagnostic emission lines (generally [O~III] $\lambda$4363) can be measured well in luminous PNe using large telescopes. Propitiously, M31 is rife with PNe; a survey by \cite{M06} (M06) uncovered 2700 PNe in M31 that are not located in streams or other anomalous zones of the galaxy. 

Previous PN abundance studies in M31 have concentrated on the bulge:  e.g., \citet{JC99} (12 bulge, 3 disk/halo) and \citet{RSM99} (30 bulge).  In contrast, many of the PNe identified by M06 lie in the outer regions of the disk where extinction is minimal. Many of the PNe seen in projection in the outer disk of M31 are sufficiently luminous that a full set of $\alpha$-element abundances can be measured (with some effort) and compared with those of PNe at smaller $R_{gal}$ and with those in the MW at similar galactocentric distance. These PNe in M31 with large and known values of  $R_{gal}$ (and low extinction) provide an opportunity to compare the co-dependencies of $\alpha$-element abundances in this population of PNe to the population of PNe near the Sun to see if the processes of $\alpha$-element abundances are similar; this is our primary goal. We note that our abundance results will be directly comparable to those for MW PNe since our data calibration and analysis methods are identical to those employed in HKB04 and H10.

The observations and reductions are summarized in \S2. In \S3, along with the derived nebular abundances, we describe the nebular models and inferred stellar properties. In \S4 we examine abundance correlations, evaluate the population membership of these 16 PNe, and discuss the implied oxygen gradient in M31. A summary appears in \S5.

\section{Observations and Reductions}
\subsection{Target Selection}

As stated, our primary goal was to measure chemical abundances in M31 PNe and, secondarily, to examine the O/H gradient in M31's disk. Therefore, an initial tactical requirement of our program was to select PNe that, a) had [OIII] $\lambda$4363 flux that would be detectable using moderately large telescopes; and b) spanned a wide range of galactocentric distances. Accordingly we turned to the compilation of PNe in M31 by M06. We selected 16 bright PNe which are all near the bright-end cutoff of the Planetary Nebula Luminosity Function (PNLF). In addition, they are located at sky-projected values of R$_{gal}$ from 5 to about 15 kpc and are not associated with star streams or satellites (Wilkinson, private communication); they are indicated in Figure~1. Assuming that these PNe lie in the disk (as we will discuss in \S4.2), this translates to rectified (in-plane) distances between 18 and 43 kpc.

\subsection{Observations}

\subsubsection{APO}
Our first step was to observe all targets at Apache Point Observatory (APO). Table~1 gives details of the observations, along with m$_{5007}$ and coordinates of the PNe. (All PNe lying within its survey footprint are visible in SDSS images in r' and g' filters.) We observed these 16 PNe between 2007 and 2010 at APO using the 3.5-m ARC telescope and Dual Imaging Spectrograph, which allows simultaneous observation from 3700-9600 \AA~with $\sim$8 \AA~resolution. All of our spectroscopic data longward of 6400\AA~comes from APO. Since the targets are point sources, we observed with slits as well matched as possible to the seeing: we used a 6$\arcmin$-long slit that was either 1.5$\arcsec$ or  2.0$\arcsec$ wide, and we binned on-chip by 2 in the spatial dimension for a 1.8 \AA~(blue) or 2.3 \AA~(red) x 0.8$\arcsec$ effective pixel size. The observing conditions over the APO full- and half-night runs varied from photometric with $\sim$1" seeing to variably cloudy with seeing $\sim$~2.5"; we lost two half-nights to high humidity. APO has no atmospheric dispersion corrector, so we attempted to position the spectrograph in real time to the parallactic angle in order to minimize such effects. Good fluxes were obtained for all of the brighter abundance diagnostic lines for H, He, N, O (except for  $\lambda$4363), Ne, and Ar; S was most often detected marginally.

\subsubsection{Gemini}
The -300 km s$^{-1}$ systemic radial velocity of M31, combined with the bright Hg~I $\lambda$4358 line from the city of Alamogordo rendered the [OIII]$\lambda$ 4363 line difficult to disentangle at APO. The $\lambda$4363 fluxes for all but two of the target PNe were obtained using Gemini North where $\lambda$4358 is invisible (see below). 

The spectral range from 3600 to 6400 \AA~was reobserved at Gemini North in October 2009 where we observed 14 of the 16 APO PNe using GMOS. PN6 and PN10 were not included in these observations, and all results for them come from the APO observations.  GMOS has ~$\sim$7\AA~resolution. We used a 5.5$\arcmin$-long slit that was 1.5$\arcsec$ wide, binning on-chip by 4 pixels in the wavelength dimension and by 2 pixels in the spatial dimension, yielding an effective pixel size of 1.8 \AA~x 0.15$\arcsec$. We were able to measure $\lambda$4363, unaffected by the Hg line, in all 14 PNe; the analogous [N~II] line, at $\lambda$5755 was detected in 13 objects. Interestingly, [O~III] temperatures derived from these higher signal-to-noise observations were almost uniformly lower (weaker $\lambda$4363) than from APO observations, confirming a well known tendency for weakly-detected lines to be systematically over-measured when accurate sky subtraction is difficult. Observing conditions at Gemini were excellent: conditions were photometric on all three nights, and seeing varied between 0.5" and 1.3." GMOS has no atmospheric dispersion corrector, so, as at APO, we positioned the spectrograph to the parallactic angle during the night. Compared with M06 our observed values of m(5007)\footnote{m(5007)$\equiv$-2.5logF(5007)-13.74 \citep{J89}} are systematically slightly fainter and agree to within $\pm$0.07 magnitudes (excluding PN6 and PN11, which were poorly centered in the spectrograph slit at APO and Gemini, respectively.)

\subsection{Data Reduction}
Data reduction was carried out in IRAF\footnote{IRAF is distributed by the National Optical Astronomy Observatory, which is operated by the Association of Universities for Research in Astronomy under cooperative agreement with the NSF}, with tasks in the {\it noao} and the {\it gemini} packages for DIS and GMOS data, respectively. The two-dimensional PN and standard star frames were bias subtracted, flat-field corrected, and wavelength calibrated. Cosmic rays were removed using  the task {\it lacosmic} \citep{vD01}. One-dimensional spectra were extracted and corrected for atmospheric extinction; finally, a standard star flux calibration was applied to the PN spectra, which were then combined into a final spectrum for each object.  Figure~2 shows an expanded Gemini spectrum of PN1 with important lines labelled; it is generally representative of all the target spectra.

\section{Results}
 
 \subsection{Line Intensities}
 
We measured emission-line fluxes from the final combined, calibrated blue and red spectra of each PN using {\it splot}. These fluxes formed the input for our abundance determinations using ELSA, our 5-level atom code \citep{J06}. The first step in the analysis is to generate a table of line intensities that have been corrected for interstellar reddening and for contamination of the hydrogen Balmer lines by coincident recombination lines of He$^{++}$. We corrected for the effects of reddening using the function of Savage \& Mathis (1979). Details of the analysis using ELSA  are described in \citet{Milingo10}. Table~2 contains the emission-line measurements. Column entries are as follows: the first column lists the ion and wavelength designation of each line; f($\lambda$) gives the value of the reddening function from Savage and Mathis (1979), normalized to H$\beta$=0; F($\lambda$) is the measured flux and estimated error, relative to H$\beta$=100; and I($\lambda$) gives the reddening-corrected intensity and estimated error, also relative to H$\beta$=100. At the bottom of each column, for each nebula we list the logarithmic reddening parameter, {\it c}(H$\beta$), the theoretical H$\alpha$/H$\beta$ ratio appropriate for the nebular temperature and density, and the log of the total observed H$\beta$ flux through the spectrograph slit.

\subsection{Plasma Diagnostics and Abundances}
 
Table~3 contains the plasma diagnostics: electron temperature, T, from [O~III] and [N~II]; and electron density, N, from [S~II]. All 16 PNe have T[O~III] calculated from $\lambda$4363. Thirteen have T[N~II] calculated from $\lambda$5755 as well; for the other three PNe, we assume T[N~II] =10,300 K.  In three PNe we were able to measure $\lambda$6312, the auroral line of [S~III], so we calculated T[S~III] in those cases (though this value is not used in the final abundance analysis). In only two PNe were both  [S~II] lines $\lambda\lambda$6717,6731 detected, enabling a direct estimate of the electron density. For most of the remaining objects we detected $\lambda$6731, but not $\lambda$6717, for which estimated upper limits imply densities in excess of 10,000~cm$^{-3}$, consistent with these being among the brightest PNe in M31. To gauge how the assumed density would affect the derived abundances, we carried out calculations for both 10000~cm$^{-3}$ and 15000~cm$^{-3}$. The resulting values differed by less than 10\%, except for N/H and N/O, which are both $\sim$20\% lower at 10000 cm$^{-3}$. Since the [S~II] lines indicate densities above this, we adopt the value of 15000~cm$^{-3}$.  

Ionic abundances derived using ELSA are given in Table~4. The first column lists the ion and wavelength (if more than one line is observed for that ion); the second column shows which temperature was used in the calculation; ``wm" indicates a weighted mean value, (weighted by raw observed flux) for the final ionic abundance to be used in further calculation. Ionization correction factors that were derived to compute total abundances are listed in Table~3; these have been calculated in ELSA as described in \citet{KH01}. It is important to reiterate that our comparisons of abundances in M31 with those in MW PNe (\S4.1) are all based on identical methodologies at all stages of the data acquisition, calibration and analysis. The total elemental abundances are shown in Table~5; the last two columns give values for the sun \citep{asplund09} and Orion \citep{esteban04}.

\subsection{Properties of the Central Stars and their Progenitors}

\subsubsection{Cloudy Models}

For each PN we calculated a model using version 8.00 of Cloudy \citep{F98}, in which we matched a selection of nebular line ratios that are diagnostic of stellar and/or nebular properties. Table~6 lists the observed and modeled values of these ratios for each PN, along with the resulting parameters, including the luminosity, temperature, and surface gravity of the central star (CSPN), and the nebular abundances. We used atmospheres from \citet{Rauch03} with log {\it g} = 6.5. All models were a spherical shell with an inner radius of 10$^{17}$ cm; except for two, noted in the table, they were truncated (i.e. matter-bounded) to match the observed intensity of the [O~II] $\lambda$3727 line. In all cases, we were able to achieve a good match to the observed abundance ratios. Nebular outer radii and implied masses are shown in the ``Model Parameters" section in Table~6. The radii range between 0.04 and 0.08 pc, averaging 0.05 pc, and the implied nebular masses fall between 0.03 and 0.21 M$_{\odot}$, averaging $\sim$0.07 M$_{\odot}$. 

The largest source of uncertainty in the models is the (unknown) abundance of carbon, a major nebular coolant.  In order to obtain agreement with the observed line intensities and nebular temperatures, the required input carbon abundances were many times solar, which is probably not realistic. Another {\it ad hoc} remedy would have been to assume that gas density is a function of radial distance from the central star. But in either case we are introducing an unconstrained property into the models. At the same time, we point out that the primary goal of the model computations is to determine the central star's effective temperature and luminosity. Each of these parameters can be constrained by recombination lines whose strengths are relatively insensitive to electron temperature. For example, the ratio of He~II $\lambda$4686 to He~I $\lambda$5876 is positively correlated with stellar effective temperature, while the nebular luminosity of either H$\alpha$ or H$\beta$ tracks the stellar luminosity. To test this claim, we chose three PN models in which the C abundance had been raised significantly in order to obtain a match with observations (PNe 5, 6, and 11). In each case we reran the model after reducing the input C abundance so that the C/O ratio was equal to the solar ratio. The resulting change in He~II $\lambda$4686 to He~I $\lambda$5876 and the luminosity of H$\alpha$ was only a few hundredths of a dex in each case.   Therefore, we are confident that in terms of stellar parameters, our models remain robust despite the adjustment of an unconstrained abundance parameter to help bring the entire set of predicted emission line strengths into agreement with those that are observed.

\subsubsection{Analysis of Central Star Properties}
Information regarding the properties of the CSPN and their progenitors
can be inferred by comparing the positions of CSPNs in a theoretical H-R diagram which also
contains evolutionary tracks of post AGB stars. From the modeling described above, the bolometric luminosity
as well as the effective temperature of each CSPN was deduced. This information appears in
Table~6 and is listed again by object in Table~7. For each object identified in the first column of Table~7
we list $\log T_{eff}$ and $\log L/L_{\odot}$ in columns 2 and 3. Subsequent columns list the fraction of stellar luminosity emitted in $\lambda$5007, the final
(CSPN) mass, $M_f$, the initial mass of the progenitor star, $M_i$, and the main sequence lifetime, $t_{ms}$.

Figure~3 is a theoretical H-R diagram, i.e., $\log L/L_{\odot}$ versus $\log T_{eff}$, in which we show
five evolutionary tracks from the post-AGB models of \citet{vw94}. Each
track is labeled with the mass of the remnant star. In addition, we show the positions of
our 16 sample objects, based upon the information in the first two columns of Table~7. We then estimated the CSPN
masses and uncertainties by comparing object positions with the model tracks in the figure.
The resulting values are those listed in column 4 of Table~7. 

In Figure~4 we plot the extinction constant {\it c}(H$\beta$) vs. core mass for our objects. The Milky Way foreground toward M31 imposes E(B-V)=0.062 \citep{schlegel98}, which translates into a contribution to {\it c}(H$\beta$) of 0.088, assuming  a ratio of {\it c}(H$\beta$)/E(B-V)=1.43 \citep{kl85}. The points in Fig.~4 have had this foreground reddening removed. The figure reveals an evident trend of higher extinction with higher core mass. \citet{JC99} find similar behavior for their sample of mostly bulge PNe in M31. They note that this correlation is expected theoretically, as a larger core mass will evolve from a higher-mass progenitor whose evolutionary time scale is more rapid and hence, whose cloud of dust-rich ejecta still surrounds it as it becomes a high-excitation PN; this is discussed in more detail in \citet{CJ99}. The linear least-squares slope of the {\it c}(H$\beta$)-core-mass relation for our data in Fig.~4 is 4.69$\pm$0.51, which is similar to values quoted by \citet{CJ99} for PNe in the SMC (5.6$\pm$0.7) and LMC (6.3$\pm$1.3). Their value for the slope in M31 is 8.5$\pm$1.6 (excluding the three most massive cores). We have read off the data carefully from the upper left panel in their Fig.~1 (also excluding the three most massive cores), and we calculate a slope of 4.75$\pm$1.95, about half of their value, and very close to that for our PNe. If nothing else, this exercise demonstrates the extreme sensitivity of the calculated slope to slight differences in the data values for this small sample. The corrected {\it c}(H$\beta$) values for our PN sample range from 0.01 to 0.36, with a mean of 0.15. This value is lower than the 0.4 in the LMC quoted by \citet{RP10}, closer to the 0.28 (i.e., E(B-V)$\sim$0.2) found by \citet{HC09} for M94, and M33 as well as the LMC.

Figure~5 shows a comparison of the mass distribution of our 16 objects with that of two large samples of CSPN and white dwarf masses taken from the literature. The CSPN mass distribution for 91 of our MW disk objects,, are shown with thin-line bars, where masses were taken from \citet{zhang93}. Thick-line bars show the spread for a sample of 247 DA white dwarfs with
 $T_{eff} > 13, 000K$ in the Palomar Green survey as tabulated by \citet{liebert05}\footnote{\citet{liebert05} 
claim that DA white dwarfs with $T_{eff} < 13, 000K$ possess convective atmospheres
which bring helium to the surface, resulting in higher inferred masses. Thus, we have followed their lead
and have ignored all white dwarfs with temperatures below 13,000K. None of our own objects falls into this
category.}. Values on the horizontal axis refer to CSPN or white dwarf mass, while those on vertical axis indicate the number of objects within each bin, where bin width is 0.01~M$_\odot$. The CSPN masses of our 16 M31 PNe are indicated with X symbols. (Note that the vertical coordinate for these objects is arbitrary and has no meaning). The average mass and standard deviation for each sample are indicated. We estimate that the
typical uncertainty in the M31 CSPN masses is $\pm0.04 M_{\odot}$ and roughly the same for the
other two samples. Therefore, values for the average mass for the three groups of objects in Fig.~5 are
consistent with one another. Interestingly, the MW disk sample comprises about 1/3 nitrogen rich
Type~I PNe, suggesting that the average CSPN mass should be a bit higher than the
average CSPN in M31. We therefore determined separate averages for the Type~I and Type~II
MW disk CSPN and found them to be identical. We conclude that the average values in the
three samples are indistinguishable. Finally, we wish to point out that although our objects were selected based upon bright [O~III] $\lambda$5007 line strengths, this does not necessarily imply that the CSPN should have systematically higher masses than those found in the two large samples displayed in Fig.~5. As we explain in \S4.2 below, L(5007) depends on the location of the CSPN on its evolutionary track as well as on the age of the nebula. In addition, the nebular metallicity and density profile are also likely to play a role in determining [O~III] luminosity.

Values for M$_i$ in column 5 of Table~7 were derived using the initial-final mass relation:
$M_i = -3.28 + 8.55M_f$, where we have taken eq. 1 in \citet{catalan08} and solved it
for $M_i$. These initial masses range between 1.70 and 2.36 $M_{\odot}$ with an average and standard
deviation of $1.92(\pm0.19)$. Column~6 of Table~7 provides an estimate of the progenitors' main sequence lifetimes, t$_{ms}$, where these values were determined using model results in Table 45
of Schaller et al. (1992). The average is $1.34(\pm.37)$ Gyr with a range of 0.66--1.8 Gyr.

Figures~6 and 7 contain plots of seven different abundance ratios, indicated in each panel, as functions of stellar mass for our final abundances for the M31 disk objects (Table~5). Error bars indicate the abundance uncertainties provided in Table~5 along with uncertainties in the progenitor mass given in Table~7. Average values are also shown for our sample (thin dot-dashed line) and that of the MW disk (thin
dotted line) \citep{KH12}. There is close agreement between the abundance averages for M31 and the
MW disk in all cases except that of N/O. The difference in N/O averages is no doubt due
to the fact that roughly 1/3 of the PNe in our MW disk sample belong to the nitrogen-rich
Type~I class, while only 1 of 16 objects (PN5) in the M31 sample is a Type~I. This is consistent with our finding, discussed below, that the initial masses implied by the CSPN masses of our objects are below the minimum threshold for hot bottom burning and its attendant nitrogen production.

We also include the model predictions by \citet[solid bold line]{karakas10} and \citet[dashed bold line]{marigo01} in Figs.~6 and 7. (Note that Marigo did not publish predictions of neon
and sulfur.) The two sets of model predictions for He/H, N/O, and O/H are in close agreement 
with each other over the stellar mass range relevant to our sample. There is also satisfactory
agreement between models and observations in the cases of He/H, N/O, and Ne/H. 

However, theory and observation are at odds in the cases of O/H, Ne/H, S/O, Ne/O, S/H, and S/O. Both Marigo's and Karakas's models were ``tuned" to the solar abundance set published by \citet{anders89}. In the cases of O/H, Ne/H, and Ne/O the theory-observation offsets can be accounted for by assuming
that the more recent solar abundances by Asplund et al. (2009) apply. However, S/H and S/O offsets cannot be explained
in this way. Rather, observed sulfur abundances in PNe have
been known for several years to exhibit levels significantly below expected values, most likely because of a
faulty ionization correction factor for sulfur (see Henry et al. 2012). Thus, the difference
between observation and theory for S/H is easily explained by this problem, while that of
S/O is related to this same problem along with the updated solar abundances. Therefore, 
we see no major unexplainable discrepancies between our observed abundances and the
predictions of models of post AGB stars.

Finally, our observations indicate that these PNe, all near the bright end of the PNLF, share numerous characteristics: in addition to the high [O~III] luminosities they were selected for, they appear similar in size, density, and some abundance characteristics like N/O and Ne/O.

\section{Discussion}

\subsection{Abundance Correlations}
Figures~8-11 display the abundance correlations in our objects along with the corresponding data for the MW PN samples from HKB04 and H10. The top plot in Fig.~8 shows log(N/O) vs. log(He/H). The M31 PNe are shown as filled circles; MW Type~I PNe are indicated by open circles, and Type~II by open triangles;  the sun symbol indicates solar values. Note the expected, positive correlation for MW Type~I PNe, suggestive of the results of hot-bottom burning in more massive progenitors producing nitrogen at the expense of oxygen. This behavior is seen in LMC PNe, but not in SMC PNe (see \cite{KH12} for a compilation of abundance correlations in the MW, M31, and the MCs). Neither the MW Type~II nor the M31 sample exhibits a similar strong trend, arguing for a Type~II classification for  the M31 objects. The only possible exception is PN5, the M31 PN with the highest N/O, which falls above the majority of the MW Type~II objects, and into the region occupied by Type~I PNe in both this plot and the bottom plot showing log N/O vs. log(O/H). The outlier at low He/H is PN6, (observed only at APO) whose helium abundance is low and uncertain. In Figs.~8-11, the correlation between any plotted element ratio, X/O, and the measured O/H value means that if O/H is underestimated, then X/O is overestimated; the opposite is the case if O/H is overestimated.

The top plot in Fig.~9 shows log(Ne/H) vs. log(O/H). Here and in the following two figures the M31 PNe are shown as solid circles, all MW PNe as smaller open circles, and solar values as a sun symbol. The well known close correlation between neon and oxygen is evident, supporting the idea that they are produced in the same locations; the slopes ($\sim$0.97) are identical. In these plots and the following plots for argon and sulfur, the M31 PNe exhibit a tighter correlation than do the MW PNe; this is likely a reflection of the relative homogeneity of the M31 sample compared with MW sample. The lack of an obvious correlation between Ne/O and O/H, as seen in the bottom figure, argues against neon production in the progenitors of these PNe \citep{KH12}. The behavior of argon, shown in the two plots of Figure~10, is similar to that of neon, as expected, albeit with more scatter, since the derived argon abundances are less accurate. 

Sulfur is shown in Fig.~11. As evidenced by the MW PNe, the sulfur-oxygen correlation is even more diffuse than that of argon-oxygen. In addition, the determination of sulfur abundances in the M31 PNe is more challenging than usual: apart from any issues concerning the ionization correction factor, the lines from both S$^{+}$ and the predominant ionization state, S$^{+2}$, are very faint, or absent, in these PNe, and for most of them, the assumed density of 15,000 cm$^{-3}$ is a lower limit. Nevertheless, it is clear that the derived sulfur abundances for PNe in M31 PNe are below solar, indicating that here, as in the MW and the Magellanic Clouds, PNe exhibit the ``sulfur anomaly," the systematic underabundance of sulfur relative to oxygen (see \S3.3.2 and Henry et al. 2012).

Finally, we briefly mention a comparison of some of our results with those of \citet{JC99} and \citet{RSM99}, noting again that they concentrated on M31 bulge PNe. The N/O and He/H values found by \citet{JC99} extend to higher values than our sample, indicating Type~I PNe; we found only one possible Type~I object. We find that the range of oxygen abundances in our sample is very similar to the range found by these authors. Also, the relation between neon and oxygen for the samples of both, plotted in Fig.~3 of \citet{JC99} overlaps very well with our Fig.~9 (top), again demonstrating the near universality of this correlation.

\subsection{Population Membership}
We now explore whether the PNe in the M06 survey are drawn from the outer regions of the thin or thick disks of M31.  First, the PNe follow the brightness of red starlight along the minor axis out to an R-band surface brightness of 24 mag arcsec$^{-2}$ where most of the light arises in the thin disk (M06).  Beyond this radius where the thick disk is readily visible (Collins et al. 2011, Fig. 9) the statistical scatter in the PN numbers becomes large.  The dispersions of PN kinematics at large radii, Å55 km s$^{-1}$ (M06), agree at least superficially with the dispersions of stars in the thick disk, 51 km s$^{-1}$ \citep{C11}. However, as M06 point out, a flare in the thin disk like that seen in H~I can enhance the velocity dispersion.

Turning to our target PNe, we noted that they are seen at projected R$_{gal}$ between 5 and 15 kpc. Using a distance to M31 of 770 kpc \citep{fm90}, 1~arcminute corresponds to 0.224 kpc.  Assuming that the observed PNe are associated with M31's disk, we can calculate the rectified R$_{gal}$  (i.e., distance from the center of M31 in the plane of M31's disk) according to:

\begin{displaymath}
R_{rectified} = (X^2 + (Y/cos~{\it i})^2)^{1/2},
\end{displaymath}
where X and Y are the distances along the major and minor axes, respectively (in angular or linear distance), and {\it i} is M31's inclination to the plane of the sky, equal to 77.7$^{\circ}$ \citep{deV58}. For each PN we estimated the radial velocity expected for its location if it were following the rotation of the disk (Figure~37 of M06). To do this, we calculated $\theta$, the azimuthal angle of the PN's location measured eastward from the major axis in the southwest: $\theta$=$tan^{-1}${\it (X/Y)}. We then calculated $\phi$, the corresponding angle in the plane of M31:  $\phi=\theta/sin$ {\it i}. We next took the difference between the observed and expected radial velocities, and divided it by the appropriate value of the velocity dispersion, $\sigma$ (Figure~34 of M06). Of these 16 PNe, three have velocities $\sim$3$\sigma$ from that expected for disk rotation and one very near the minor axis differs by 5$\sigma$; see Table~8. Given that M06 note substantial velocity variations all along the major axis we conclude that these objects are likely members of M31's disk  population. In that case, their rectified R$_{gal}$ values range from 18 to 43 kpc.  For comparison, the window of $R_{gal}$ for H10's MW PN sample lies between 5 and 20 kpc.

Another potential discriminator of PN-stellar-population membership is population age.  However, the [O~III] luminosity criterion that we used to select our targets is likely to select only those PNe with the properties that we derived; i.e., final core masses between 0.58 and 0.7~M$_{\odot}$ no matter what the population of stars from which they originate.  This is shown nicely in the results of an extensive theoretical study portrayed in Figs.~3 and 4 of \citet{Mendez}.  Their models show that CSPNe with masses less than $\sim$0.58~M$_{\odot}$ never form PNe, a result that is empirically confirmed in our Fig.~5 for PNe in the MW and M31.  Above 0.7~M$_{\odot}$ the [O~III] peak is so brief ($\sim$100y) that the corresponding CSPNe are highly unlikely to be represented in any limited sample.  WhatÕs more, the fraction of massive progenitor stars that form such massive CSPNe is small, and their initial masses so large that such stars would dwell in the range of R$_{gal}$ where H~II regions are also plentiful.

There is additional, albeit circumstantial, evidence that our population of PNe is associated with the thin disk.  Our PN age estimates can be compared to population ages measured from C-M diagrams in the same vicinity. We are unaware of stellar age analyses at the outer age of the thin disk where our PNe are sampled.  However, the ages of the PNe are substantially younger than the ages of the stellar thick disk, 6 -- 8 Gy, or the halo (\citet{hammer10} and references therein).  The mass function of such a population will not produce CSPNe above the mass floor of 0.58~M$_{\odot}$.  Thus, by inference, our PNe are in the outer thin disk where all of the H~I is located  (\citet{Braun}; \citet{Chemin}).

Their metallicities also suggest an association of our target PNe and the thin disk. Fig.~12 indicates that the metallicity of the PNe beyond R$_{25}$ is only slightly below solar. Ignoring [O/Fe] gradients, this does not agree particularly well with the metallicity of thin-disk stars at about the same R$_{gal}$, [Fe/H] $\approx$-0.7. The disagreement is even worse for [Fe/H] in the thick disk, -1.0 (\citet{C11} and references therein). However, the comparison of O/H with Fe/H is complex since [O/Fe] also evolves \citep{Chiappini09} at rates that depend on historical supernova activity, gas infall, and other factors discussed by \citet{C11}.

Finally, the radial O/H gradient of PNe matches well with that of H~II regions which presumably form exclusively in the thin disk (see \S4.3 below and Fig.~12). This is probably no accident. All considered, it seems much more likely that the PNe formed and evolved in the outer parts of the thin disk than in the halo or thick disk.

\subsection{The Oxygen Gradient}
Figure~12 shows the M31 oxygen gradient as exhibited by this sample of PNe; also plotted are several M31 population~I abundance indicators taken from the literature and listed in the legend. For comparison we include PN data for the MW oxygen gradient taken from H10 and references therein. H10 derived a slope of -0.058$\pm$0.006 dex/kpc, while \citet{SH10} found a value of -0.023$\pm$0.006. Note that we have plotted the abundances as a function of $R_{25}$ \citep{gg98} to account for differences in disk scale length. Using a simple (unweighted) regression, we find a value for the M31 PN gradient of -0.011$\pm$0.004 dex/kpc. For comparison, \citet{B10} used a weighted regression to derive a PN oxygen gradient in M33 of -0.013$\pm$0.016 dex/kpc for a sample double the size of ours, and \citet{SMVG10} calculated a linear slope of -0.055$\pm$0.02 dex/kpc for PNe in M81.

It is clear from Fig.~12 that the dispersion in O/H values around the mean M31 PN gradient is smaller than the scatter found in local MW PNe. In the MW, distance uncertainties may account for some of the scatter. We emphasize again that all of the PN data shown in Fig.~12 are comparable in quality, the data calibration and reddening corrections are all similar, and O/H abundances were measured from the reddening-corrected line ratios in the same way. Among H10's results regarding the MW oxygen gradient is the conclusion that as much as 40\% of the scatter about the mean gradient for MW PNe may be intrinsic; i.e., due to real differences among PN oxygen abundances  at the same R$_{gal}$, perhaps due to hot-bottom burning (HBB); see e.g., \citet{karakas10}.  But this should not be a factor for the current sample of M31 PNe, all of whose progenitors have masses below the HBB threshold ($\sim$~3 M$_{\odot}$). Future observations of PNe spanning an even wider range of R$_{gal}$ will offer a deeper look into the behavior of M31's oxygen gradient.

\section {Summary and Conclusions}
We have obtained high-quality spectra for 16 PNe in the disk of M31 whose projected galactocentric distances range between 5 kpc and 15 kpc, corresponding to rectified distances from 18 kpc to 43 kpc. Their spectra contain measurable and clean [O~III] $\lambda$4363 allowing direct temperature determinations crucial for accurate abundance determinations. The oxygen gradient implied by our PN abundances and galactocentric distances is similar to that found for PNe in M33 by \citet{B10}, but shallower than the MW PN gradient fit of H10 and the M81 result of \citet{SMVG10}.

We have calculated He/H, O/H, N/O, Ne/O, Ar/O and S/O ratios for these PNe and compared them with analogous values for MW disk PNe. The M31 PNe are found to display the same correlations and general behaviors as Type~II MW PNe. One exception may be PN~5, whose abundance profile and inferred progenitor mass indicate that it is a possible Type~I. 

Photoionization models using our derived abundances yield central star properties indicating that these PNe originated from stars with main sequence masses between 1.7 and 2.4 $M_{\odot}$ and ages younger than 2 Gyr. Together with their kinematics, this indicates that they belong to M31's extended thin disk. The distribution of core masses in our sample appears indistinguishable from the distribution of MW PN core masses and from the \citet{liebert05} white dwarf mass distribution. We examined various abundance ratios as a function of progenitor mass and compared them with theoretical predictions. After accounting for the effects of updated solar abundances in the models we find reasonable agreement for all cases, save S/H and S/O. The PN sulfur deficit has long been intractable, but may be yielding to new efforts at correcting it \citep{henry12}.

The absence of many Type~I PNe in our sample is expected since there is no sign of the formation of the massive progenitor stars in the ambient stellar population. These PNe lie beyond the spiral arms where most massive stars form as evidenced by the paucity of H~II regions with R$_{gal}$ between 15 and 45 kpc. The computed mass range of the CSPN progenitors
of 1.70 to 2.36 $M_{\odot}$ places them well below the threshold for hot-bottom burning and significant
nitrogen production during third dredge-up. This is consistent with what we see in
the observed abundance pattern, in which 15 of our 16 PNe are non-Type I and
have roughly solar N/O ratios near 0.17 \citet{asplund09}. It is also likely that these
stars will prove to have solar C/O values, as their mass range is still a bit too low for carbon
to be significantly enhanced during 3rd dredge-up, according to Karakas's results.

This work demonstrates the utility of PNe for exploring abundances outside of and beyond the spiral arms of galaxy disks, particularly in M31. Bright PNe are not only available in the inner portion of the disk but also out to large galactocentric distances, where the density of H~II regions is low. Though the identification of H~II regions in M31 has recently been augmented by \citet{azimlu11} whose catalog contains more than 3600 H~II regions extending out to $\sim$24 kpc, these authors confirm that M31 is lacking in the high-luminosity H~II regions that would be most amenable to abundance studies. Therefore, PNe continue to offer the best opportunity for evaluating the chemical evolution of the interstellar medium of M31 across its disk. We have explored the spectroscopic properties of a sample of PNe near the bright-end cutoff of the PNLF. M06 showed that the PNLF is universal throughout M31, so that what we learn from this sample may also apply to similar PNe in other parts of M31 and more universally where PNe are forming and evolving.

\section{Acknowledgments}
KBK, BB, and RBCH are grateful to our institutions and to the NSF for support under grants AST-0806490, AST-088201, AST-0806577, respectively. We also thank the staffs at APO and NOAO/Gemini for their invaluable assistance. KBK and EMML acknowledge NOAO for generous travel support. The authors thank Williams College students Matthew Hosek, Aven King, and Alice Sady for their work on the Cloudy modeling. Finally, we thank the anonymous referee for a thorough manuscript reading and several suggestions that have improved the paper.

\clearpage




\begin{figure}
\epsscale{.80}
\plotone{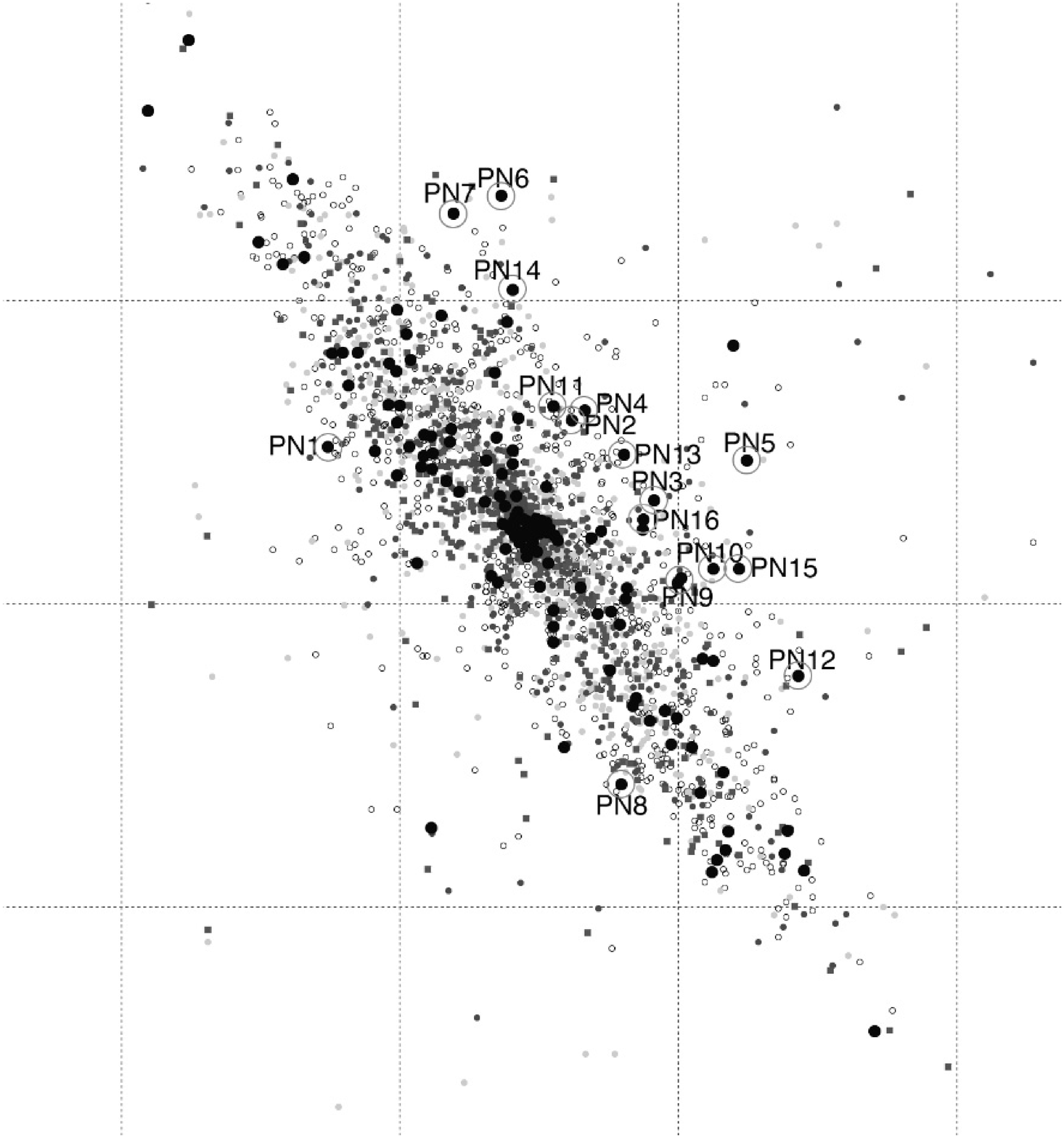}
\caption{M31 PNe observed by Merrett et al. (2006) with our target PNe labeled and circled. Dot size indicates relative brightness. Central coordinates of M31 are 00h 42m 44.3s, +41$^{\circ}$ 16' 08.5"; horizontal divisions are 5m; vertical divisions are 1$^{\circ}$.}
\end{figure}


\begin{figure}
\epsscale{0.8}
\plotone{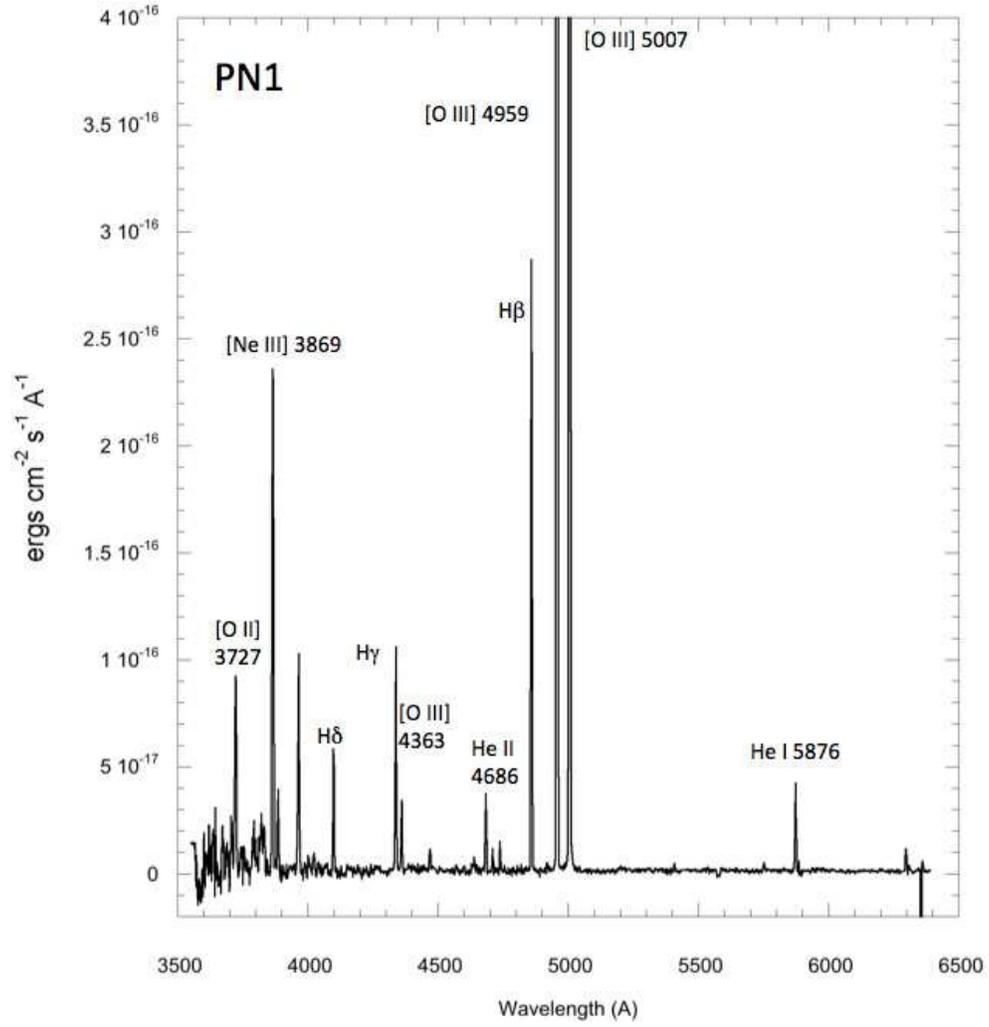}


   \caption{Gemini spectrum (1-hr integration) of PN1, with important lines labeled.}
\end{figure}

\begin{figure}
\begin{center}
\includegraphics[scale=0.6,angle=270]{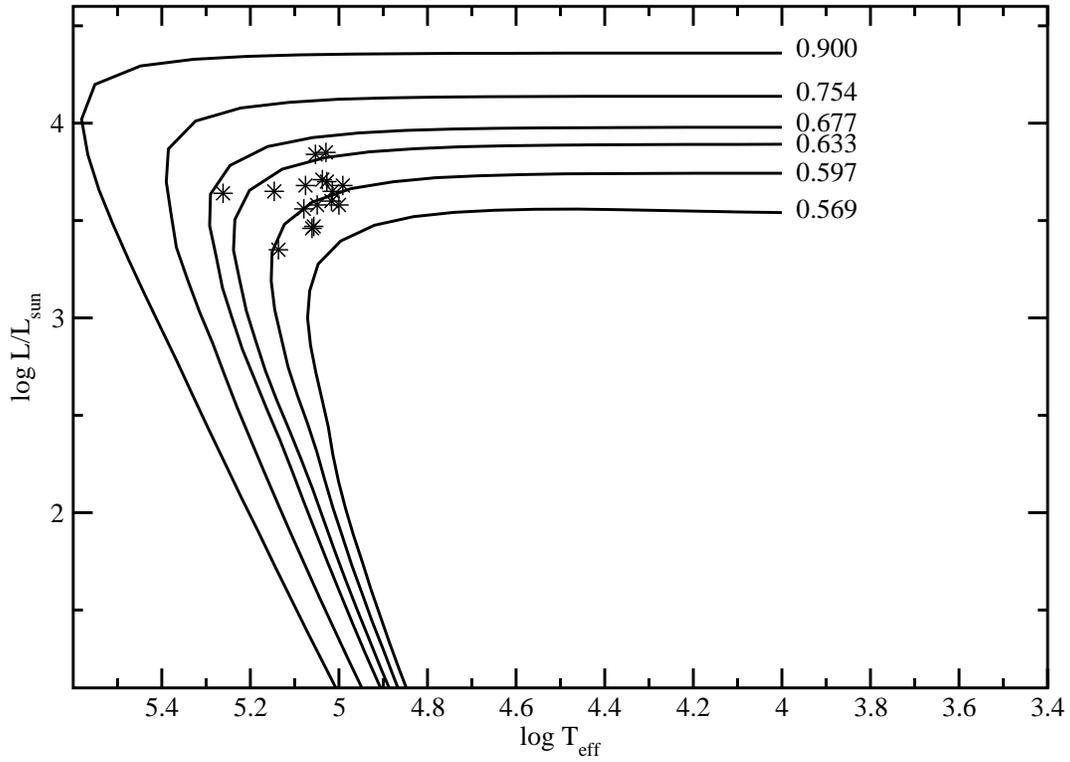}
\caption{Plot of $\log L/L_{\odot}$ versus $\log T_{eff}$, showing the positions of our 16 central stars listed in Table~8 relative to the model tracks from \citet{vw94}, where the final mass, $M_f$, corresponding with each track is indicated.}
\end{center}
\label{vw94_tracks}
\end{figure}

\begin{figure}
\begin{center}
\epsscale{0.8}
\plotone{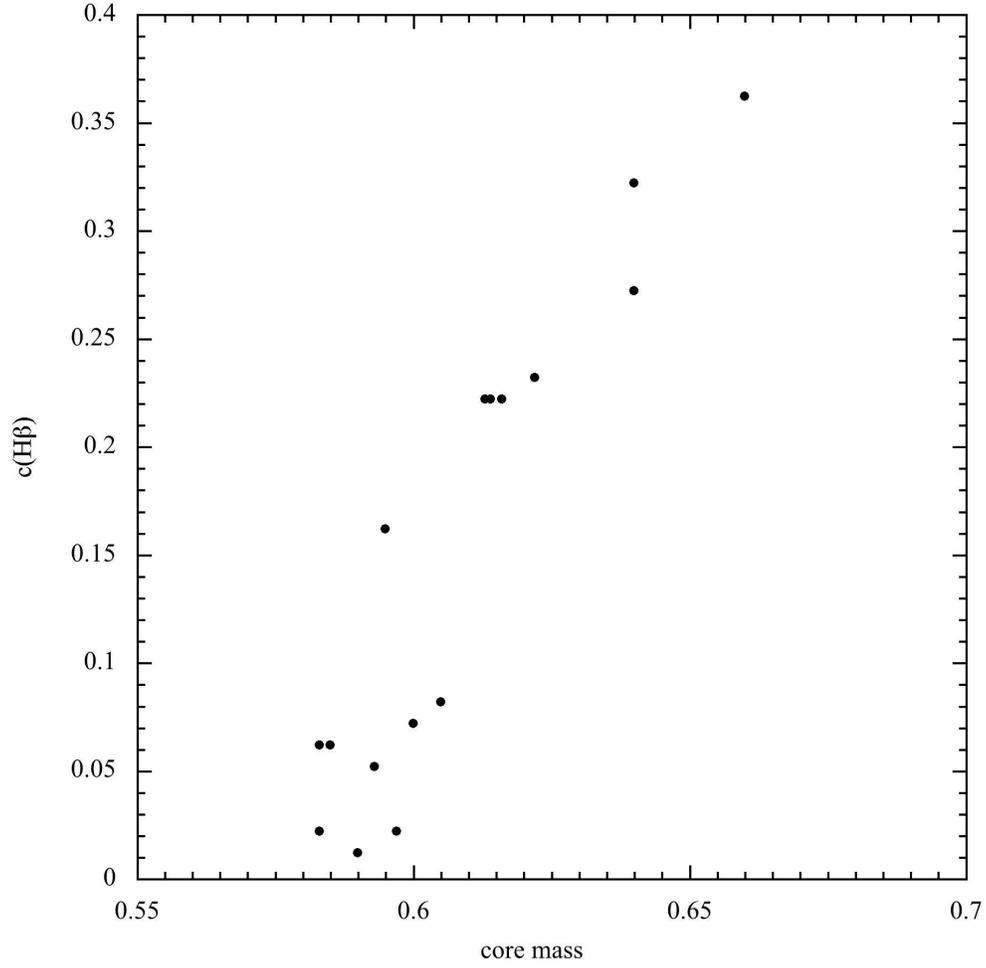}
\caption{Extinction constant {\it c} vs. core mass for our sample PNe. Foreground Galactic extinction has been subtracted off (see text).}
\end{center}
\end{figure}

\begin{figure}
\begin{center}
\includegraphics[scale=0.6,angle=270]{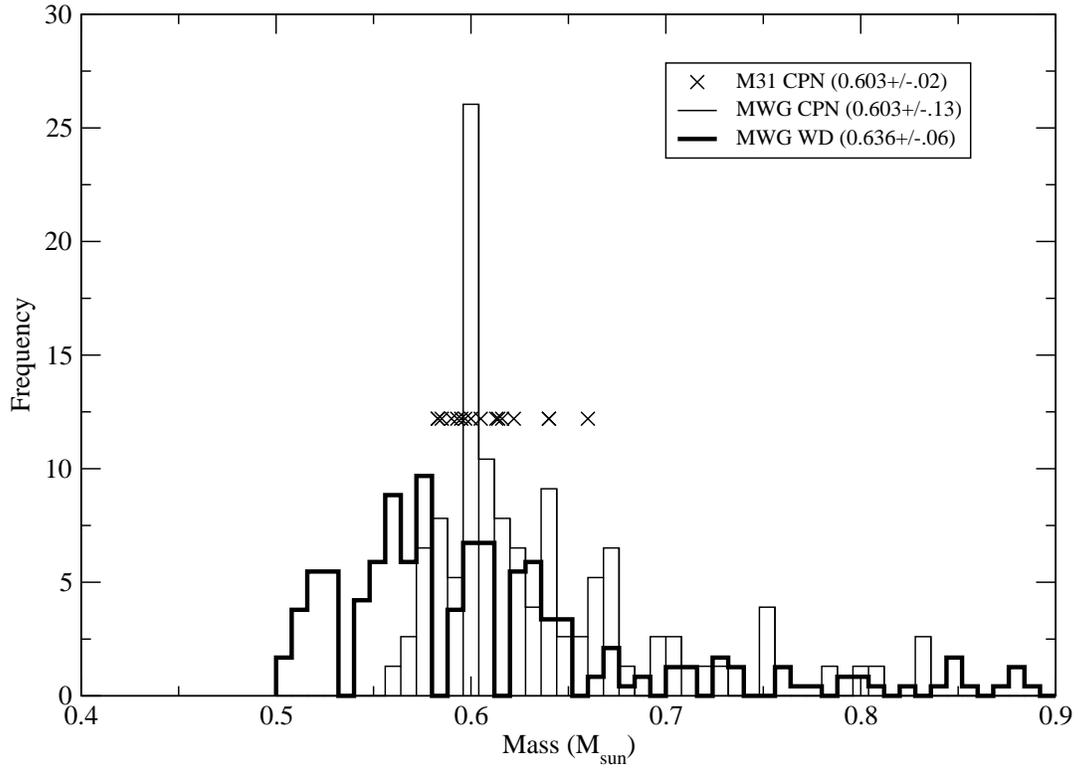}
\caption{Mass distribution of 297 white dwarf stars in the MW disk from the Palomar Green survey presented by \citet[thick-line bars]{liebert05}; 91 MW disk CSPN from \citet[thin-line bars]{zhang93}; and CSPN masses of our 16 M31 PNe (each marked by an 'X') . Averages and standard deviations are listed in the legend for each group of objects.}
\end{center}
\label{mass_hist}
\end{figure}

\begin{figure}
\begin{center}
\includegraphics[scale=0.6,angle=270]{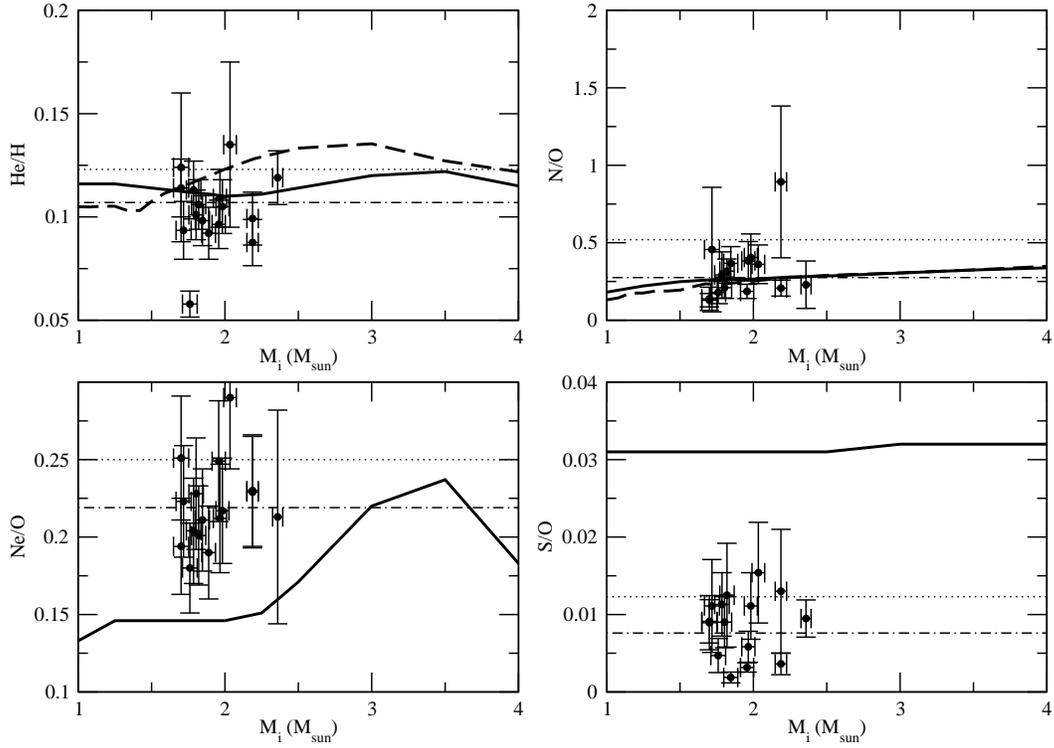}
\caption{Observed number abundance ratios of He/H, N/O, Ne/O, and S/O (as indicated
on the vertical axes) versus progenitor mass, M$_i$, in solar masses for our sample of 16 PNe in
the disk of M31. Sample averages are given by dot-dashed lines, while the average values
from our MW disk sample is shown with a dotted line. Post-AGB model predictions by
\citet[bold dashed lines]{marigo01} and \citet[bold solid lines]{karakas10} are also shown.}
\end{center}
\label{m31_karakas1}
\end{figure}

\begin{figure}
\begin{center}
\includegraphics[scale=0.6,angle=270]{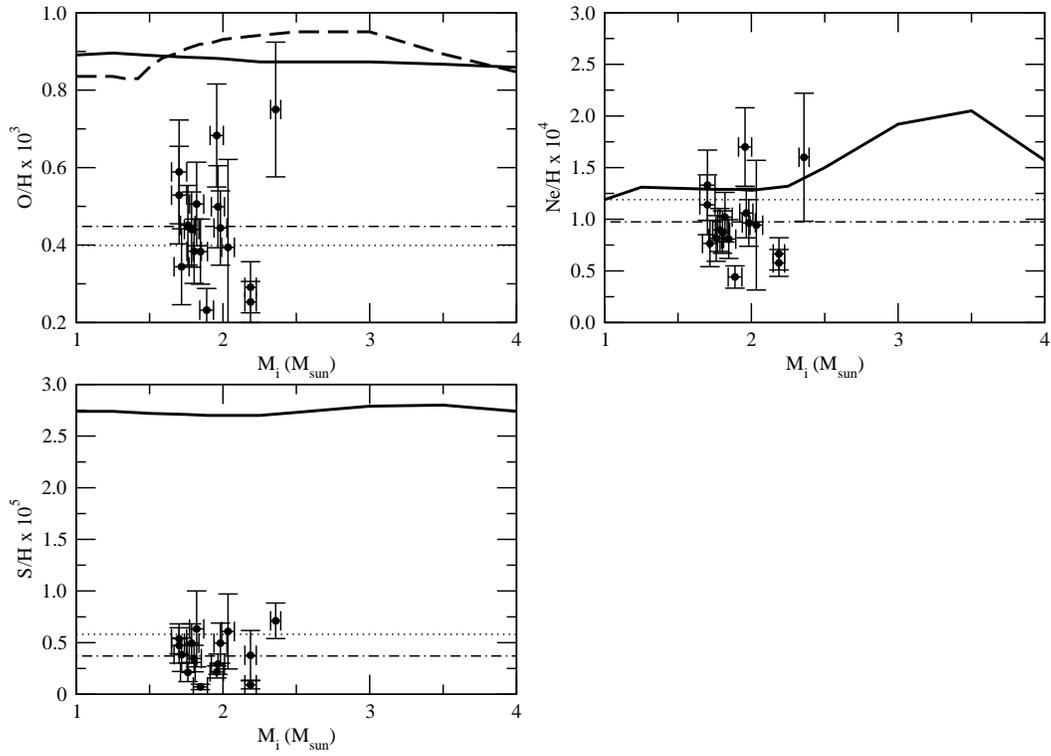}
\caption{Same as Fig.~6 but for O/H, Ne/H, and S/H.}
\end{center}
\label{m31_karakas2}
\end{figure}

\begin{figure}
\begin{center}
\epsscale{1.25}
\plottwo{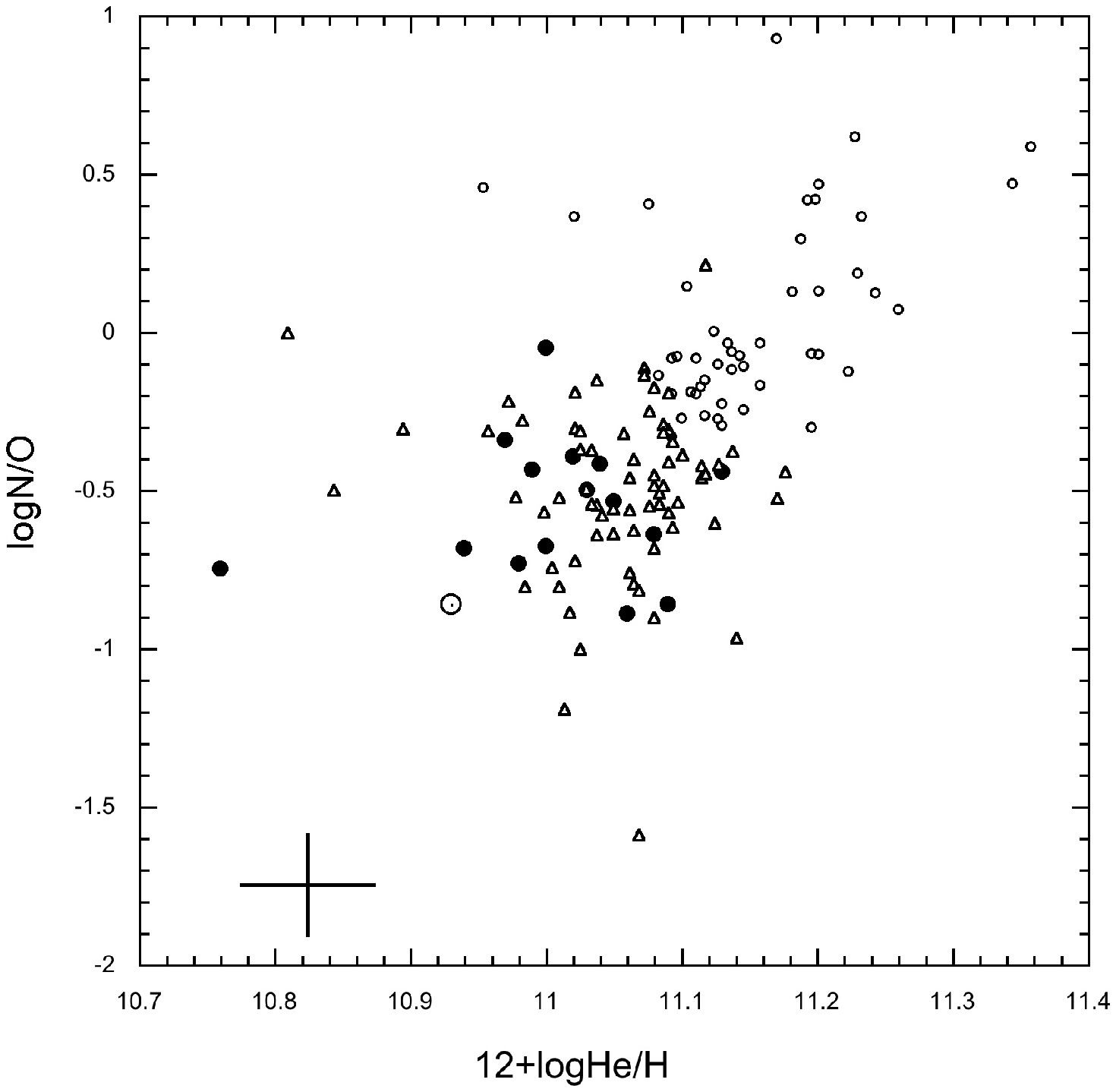}{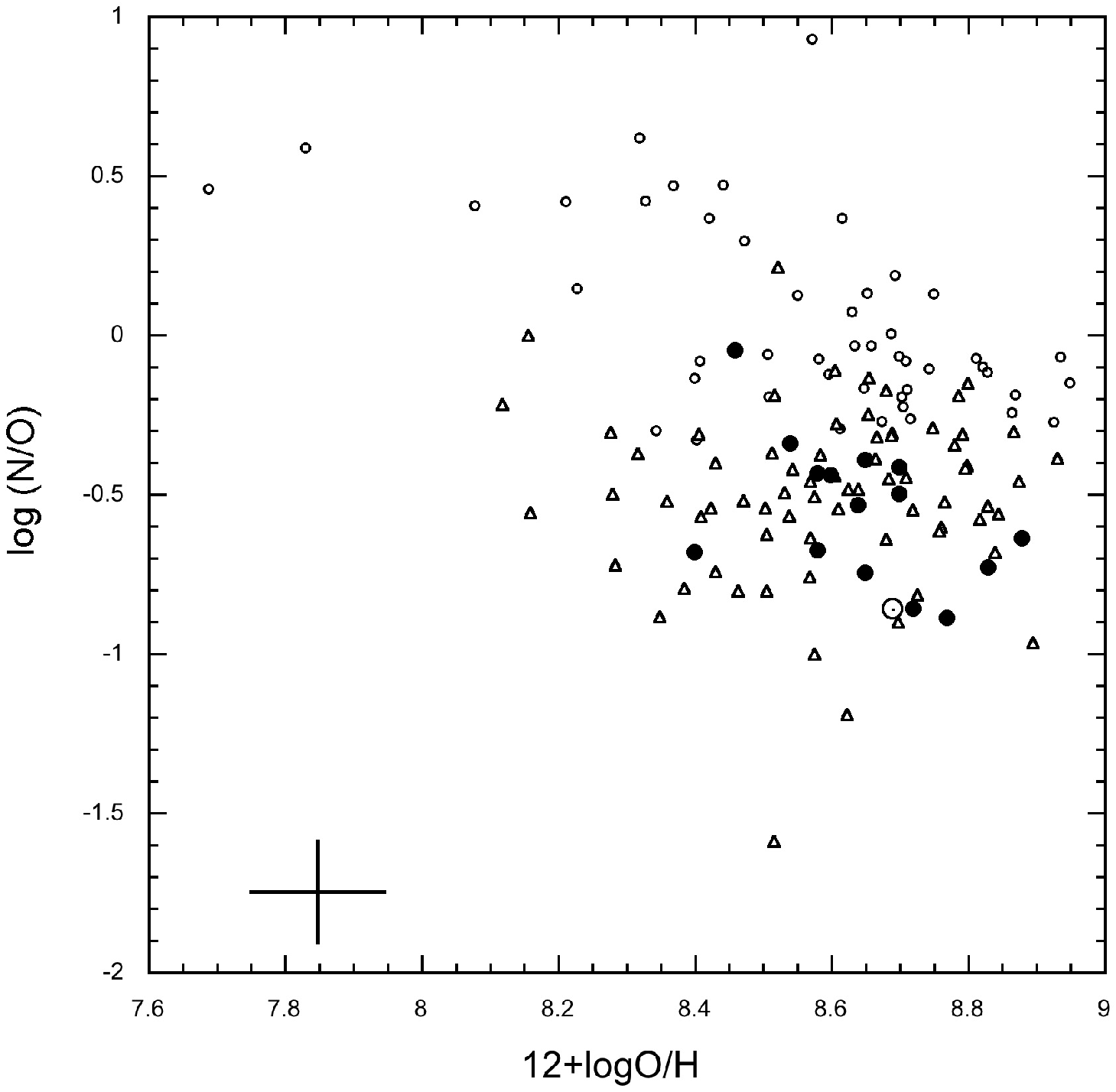}
\caption{{\it top}: log(N/O) vs. log(He/H) for our sample of M31 PNe (filled circles), Milky Way disk PNe from HKB04 and H10 (Type~I: small open circles; Type~II: small open triangles), and the Sun from \citet{asplund09} (sun symbol); {\it bottom}: log(N/O) vs. log(O/H). Representative error bars are shown in all plots; in addition, for all log(X/O)-log(O/H) plots, the correlation between the two values means that underestimating O/H increases X/O and overestimating O/H decreases X/O.}
\end{center}
\end{figure}

\begin{figure}
\begin{center}
\epsscale{1.25}
\plottwo{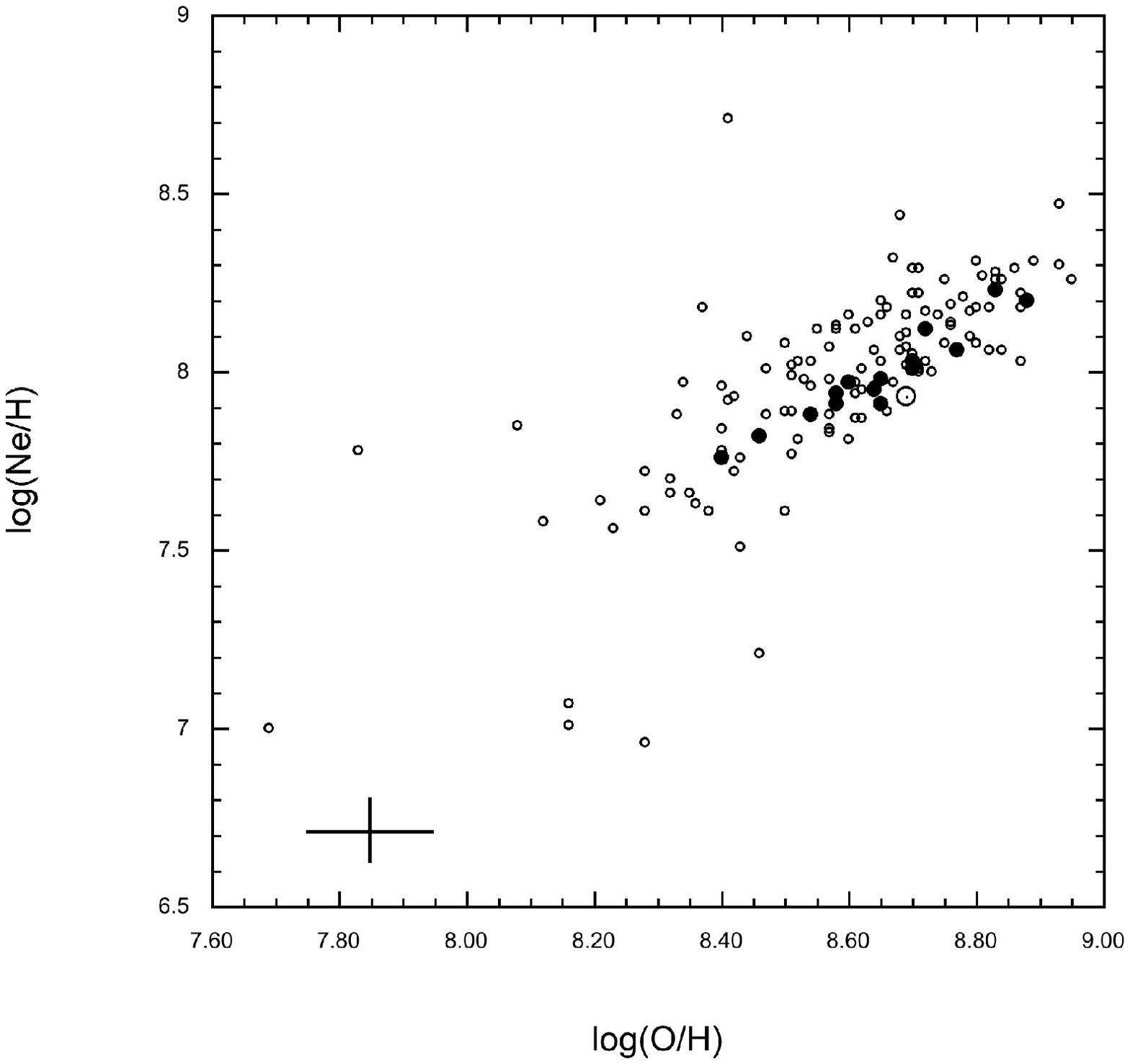}{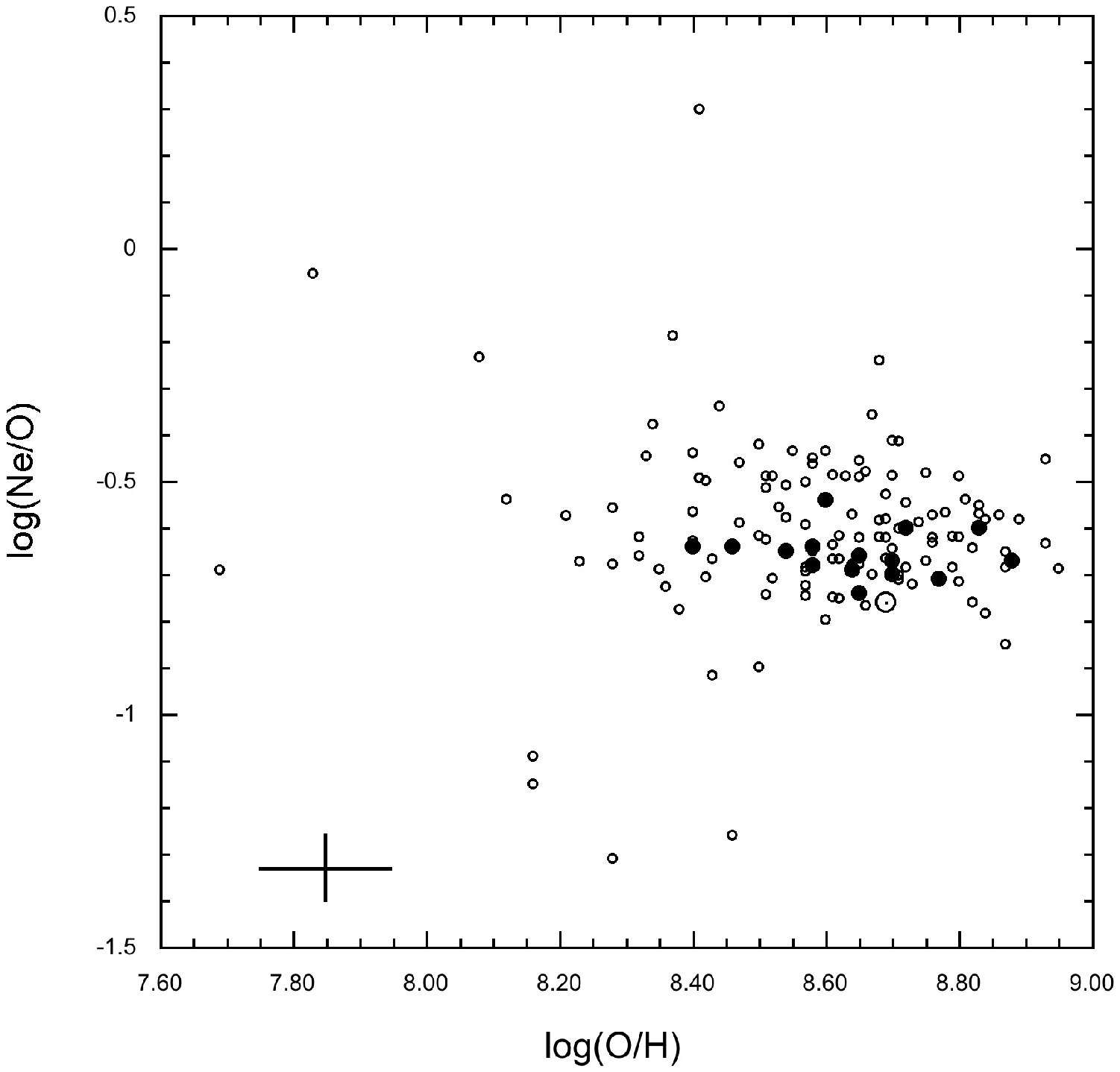}
\caption{{\it top}: log(Ne/H) vs. log(O/H); {\it bottom}: log(Ne/O) vs. log(O/H). M31 PNe are shown as solid circles, Milky Way disk PNe as small open circles, and the Sun as a solar symbol. }
\end{center}
\end{figure}

\begin{figure}
\begin{center}
\epsscale{1.5}
\plottwo{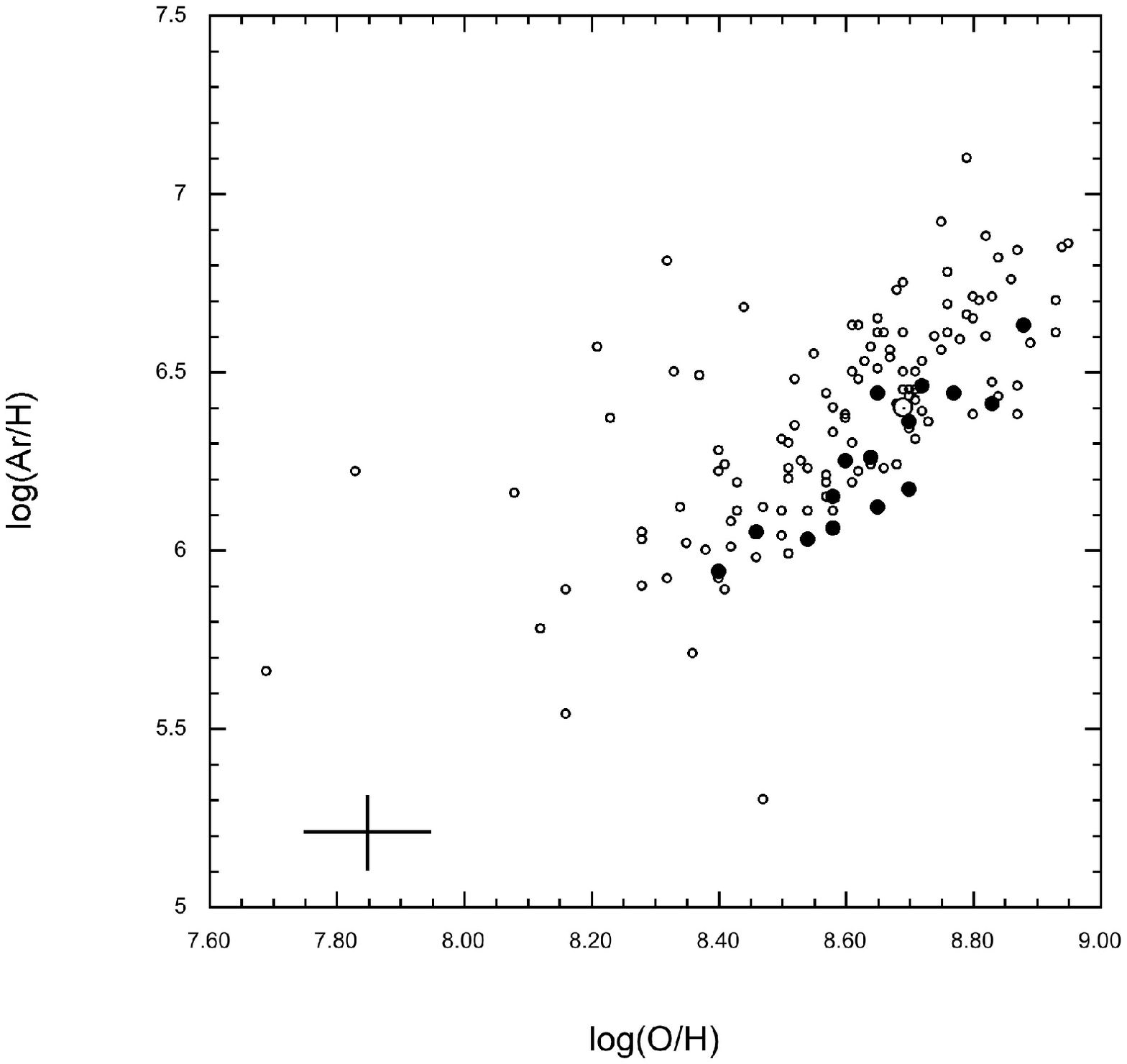}{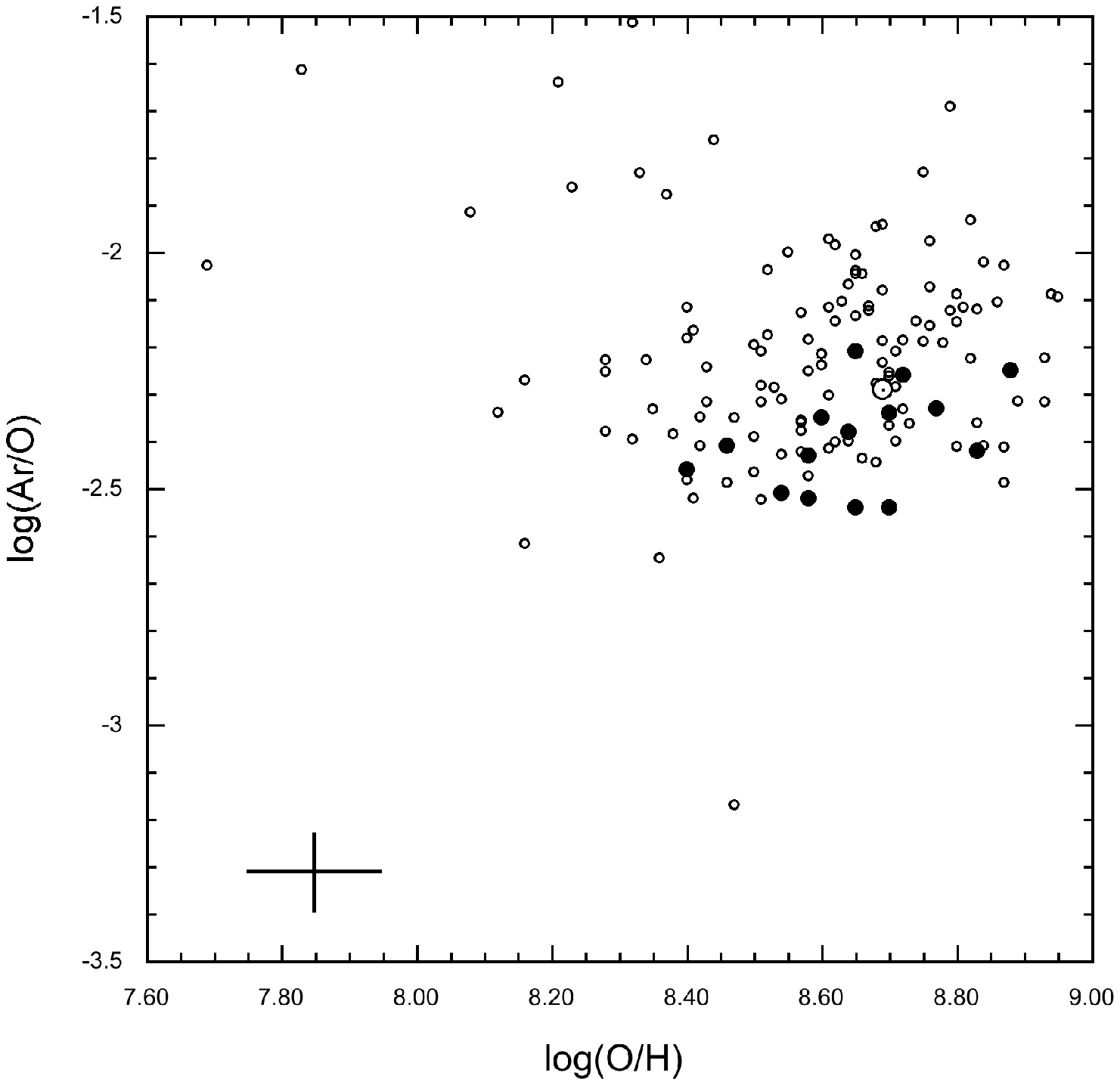}
\caption{{\it top}: log(Ar/H) vs. log(O/H); {\it bottom}: log(Ar/O) vs. log(O/H). Symbols as in Fig.~9.}
\end{center}
\end{figure}

\begin{figure}
\begin{center}
\epsscale{1.25}
\plottwo{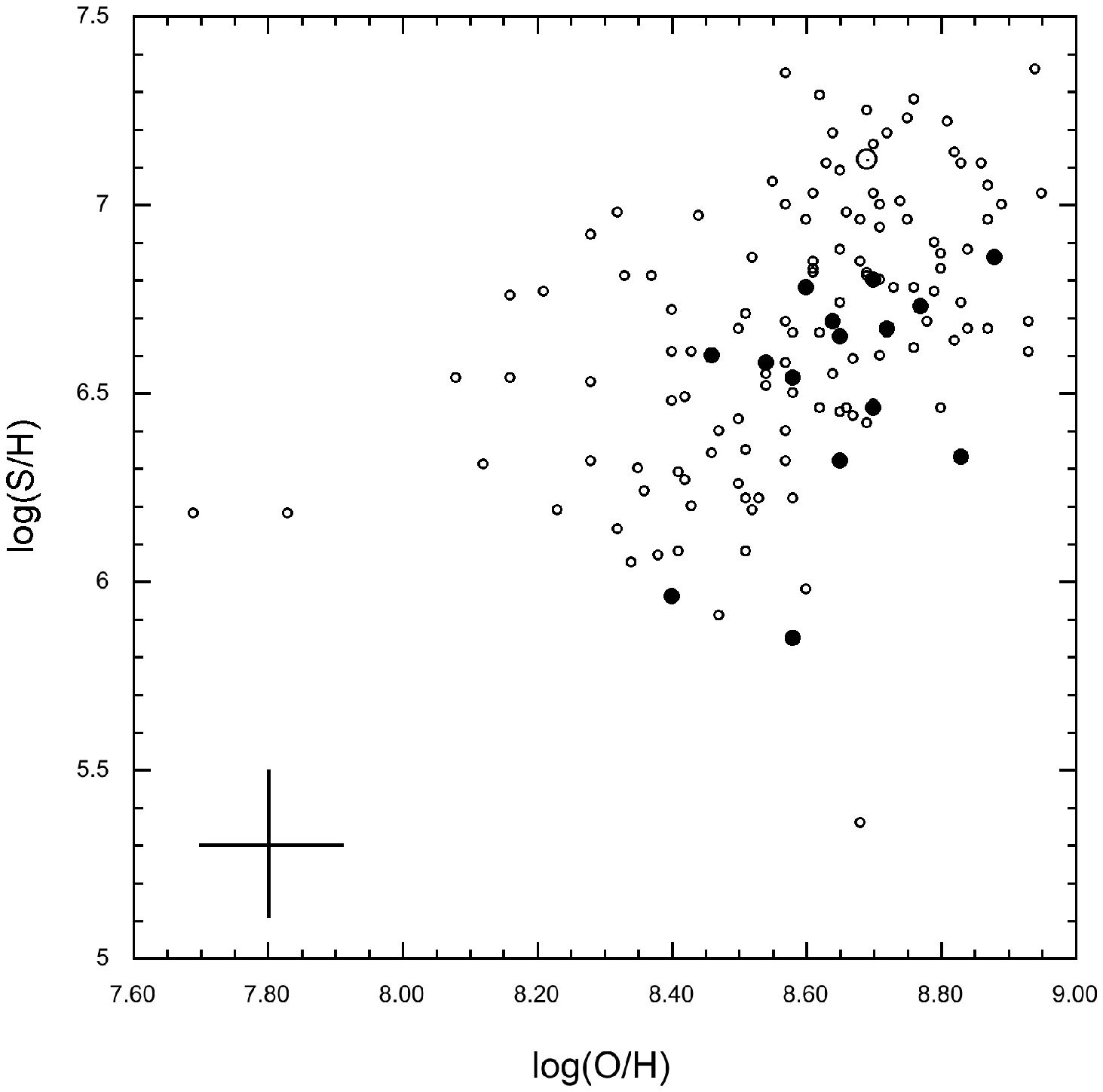}{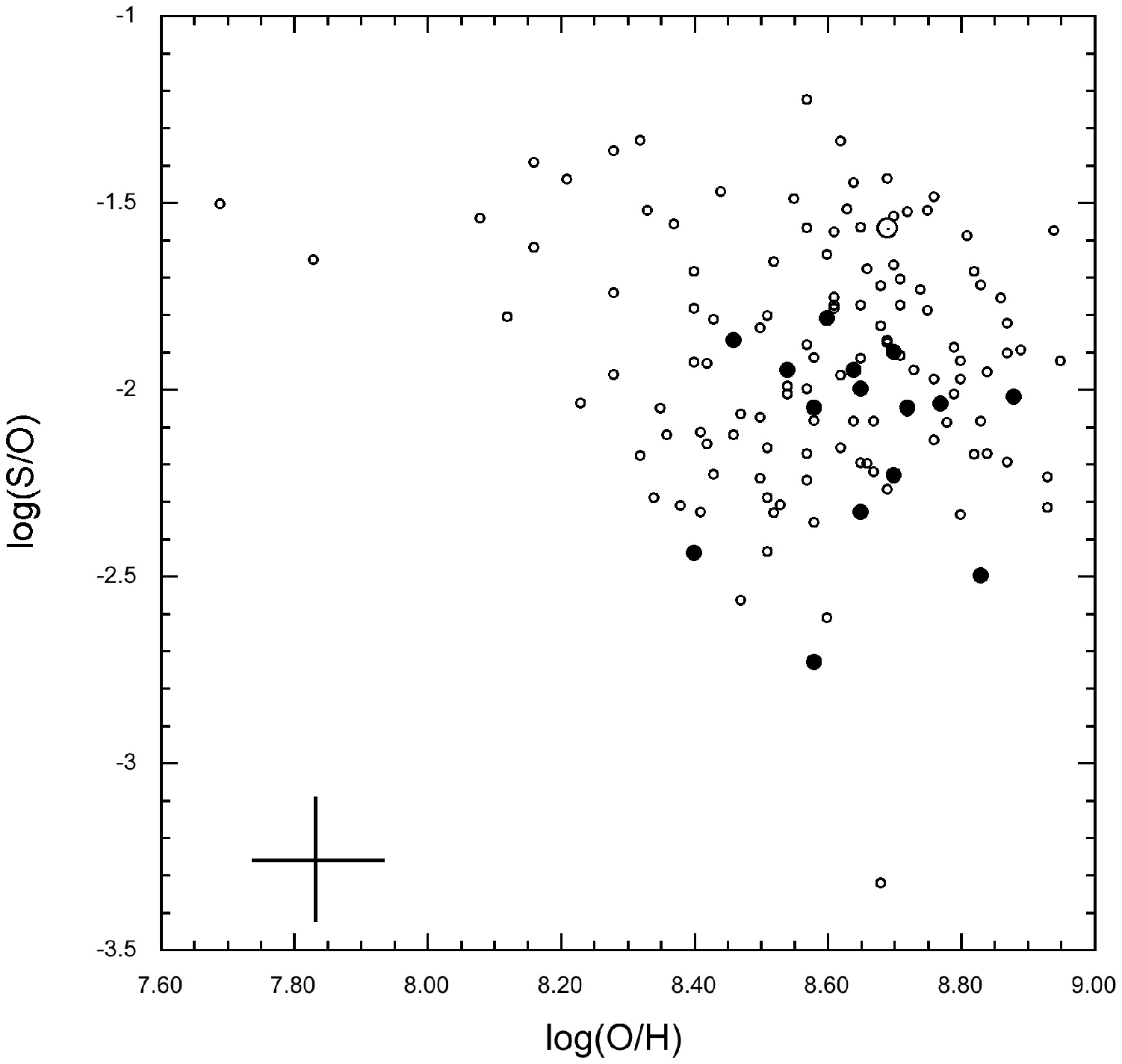}
\caption{{\it top}: log(S/H) vs. log(O/H); {\it bottom}: log(S/O) vs. log(O/H). Symbols as in Fig.~9.}
\end{center}
\end{figure}

\begin{figure}
\begin{center}
\includegraphics[scale=.6,angle=-90]{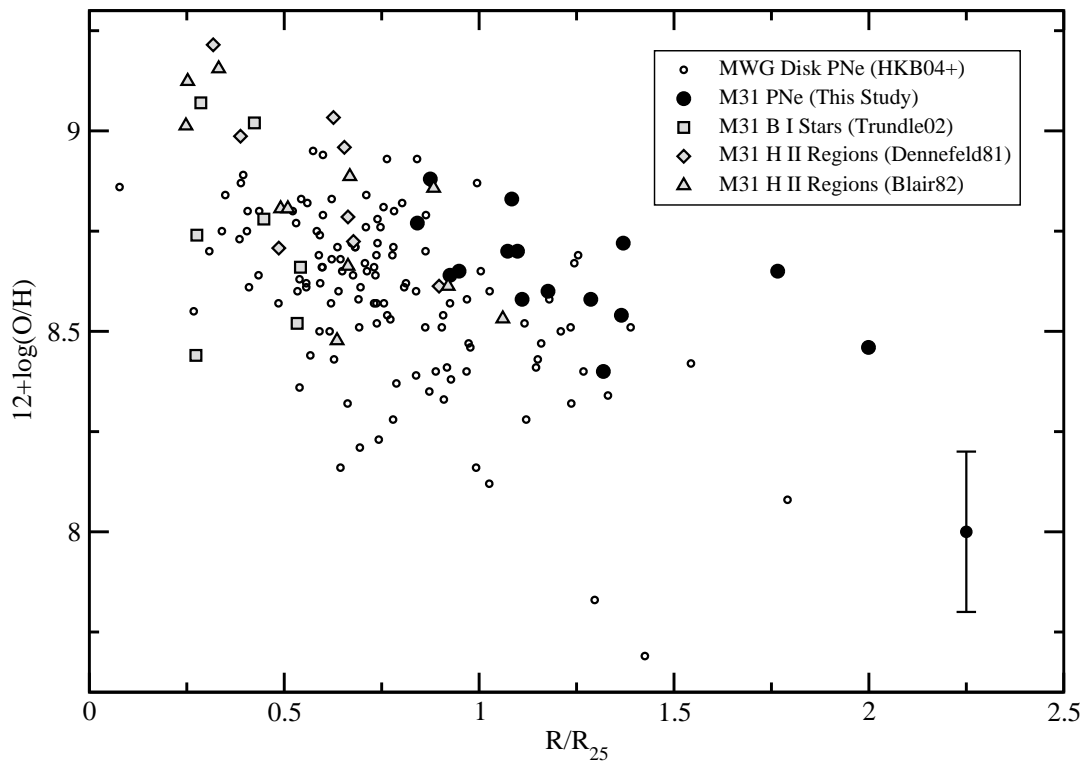}
\caption{The oxygen gradient in M31 PNe. R$_{25}$ values for M31 (22.4 kpc) and the Milky Way (13.4 kpc) were taken from \citet{gg98}. M31 PNe are shown as filled circles; other M31 oxygen probes are shown as well; sources are given in the legend. We include Milky Way PNe (small circles) from HKB04, H10 and references therein; the R$_{gal}$  window for these MW  PNe lies between 5 and 20 kpc. A representative abundance uncertainty ($\pm$0.2 dex) is indicated in the lower right.}
\end{center}
\end{figure}

\clearpage



\end{document}